\documentstyle[12pt,epsf]{article}

\expandafter\ifx\csname amssym12.def\endcsname\relax \else\endinput\fi
\expandafter\edef\csname amssym12.def\endcsname{%
       \catcode`\noexpand\@=\the\catcode`\@\space}
\catcode`\@=11
\def\undefine#1{\let#1\undefined}
\def\newsymbol#1#2#3#4#5{\let\next@\relax
 \ifnum#2=\@ne\let\next@\msafam@\else
 \ifnum#2=\tw@\let\next@\msbfam@\fi\fi
 \mathchardef#1="#3\next@#4#5}
\def\mathhexbox@#1#2#3{\relax
 \ifmmode\mathpalette{}{\m@th\mathchar"#1#2#3}%
 \else\leavevmode\hbox{$\m@th\mathchar"#1#2#3$}\fi}
\def\hexnumber@#1{\ifcase#1 0\or 1\or 2\or 3\or 4\or 5\or 6\or 7\or 8\or
 9\or A\or B\or C\or D\or E\or F\fi}

\font\tenmsa=msam10 scaled\magstep1
\font\sevenmsa=msam7 scaled\magstep1
\font\fivemsa=msam5 scaled\magstep1
\newfam\msafam
\textfont\msafam=\tenmsa
\scriptfont\msafam=\sevenmsa
\scriptscriptfont\msafam=\fivemsa
\edef\msafam@{\hexnumber@\msafam}
\mathchardef\dabar@"0\msafam@39
\def\dashrightarrow{\mathrel{\dabar@\dabar@\mathchar"0\msafam@4B}}
\def\dashleftarrow{\mathrel{\mathchar"0\msafam@4C\dabar@\dabar@}}

\def\ulcorner{\delimiter"4\msafam@70\msafam@70 }
\def\urcorner{\delimiter"5\msafam@71\msafam@71 }
\def\llcorner{\delimiter"4\msafam@78\msafam@78 }
\def\lrcorner{\delimiter"5\msafam@79\msafam@79 }
\def\yen{{\mathhexbox@\msafam@55 }}
\def\checkmark{{\mathhexbox@\msafam@58 }}
\def\circledR{{\mathhexbox@\msafam@72 }}
\def\maltese{{\mathhexbox@\msafam@7A }}

\font\tenmsb=msbm10 scaled\magstep1
\font\sevenmsb=msbm7 scaled\magstep1
\font\fivemsb=msbm5 scaled\magstep1
\newfam\msbfam
\textfont\msbfam=\tenmsb
\scriptfont\msbfam=\sevenmsb
\scriptscriptfont\msbfam=\fivemsb
\edef\msbfam@{\hexnumber@\msbfam}

\def\widehat#1{\setbox\z@\hbox{$\m@th#1$}%
 \ifdim\wd\z@>\tw@ em\mathaccent"0\msbfam@5B{#1}%
 \else\mathaccent"0362{#1}\fi}
\def\widetilde#1{\setbox\z@\hbox{$\m@th#1$}%
 \ifdim\wd\z@>\tw@ em\mathaccent"0\msbfam@5D{#1}%
 \else\mathaccent"0365{#1}\fi}
\font\teneufm=eufm10 scaled\magstep1
\font\seveneufm=eufm7 scaled\magstep1
\font\fiveeufm=eufm5 scaled\magstep1
\newfam\eufmfam
\textfont\eufmfam=\teneufm
\scriptfont\eufmfam=\seveneufm
\scriptscriptfont\eufmfam=\fiveeufm

\csname amssym.def\endcsname

\newif{\ifcomentarios}
\comentariosfalse
\renewcommand{\theequation}{\thesection.\arabic{equation}}
\newcommand{\be}{\begin{equation}}
\newcommand{\ee}{\end{equation}}
\newcommand{\bma}{\begin{displaymath}}
\newcommand{\ema}{\end{displaymath}}
\newcommand{\bc}{\begin{center}}
\newcommand{\ec}{\end{center}}

\newcommand{\text}{\rm}

\newcommand{\uflex}
{{\scriptstyle {\raise 9pt\hbox{$\backslash$}\,\!\!\!\!\!\Bigg\vert}}}

\newcommand{\ncm}{\newcommand}
\ncm{\rncm}{\renewcommand}
\ncm{\id}{{\bf 1}}
\ncm{\beq}{\begin{equation}}
\ncm{\eeq}{\end{equation}}
\ncm{\ba}{\begin{array}}
\ncm{\bea}{\begin{eqnarray}}
\ncm{\beanon}{\begin{eqnarray*}}
\ncm{\ea}{\end{array}}
\ncm{\eea}{\end{eqnarray}}
\ncm{\eeanon}{\end{eqnarray*}}
\ncm{\fns}{\footnotesize}
\ncm{\setc}[1]{\setcounter{equation}{#1}}
\newcounter{eqnr}
\newenvironment{eqnarrayabc}{\stepcounter{equation}
  \setcounter{eqnr}{\value{equation}}\setc{0}
  \rncm{\theequation}{\thesection.\arabic{eqnr}\alph{equation}}
  \begin{eqnarray}}{\end{eqnarray}\setc{\value{eqnr}}}
\ncm{\eqboxabc}[3]{\newline\parbox[t]{1.5cm}{#1}\hfill
  \parbox[b]{12cm}{\begin{eqnarray*} #3\end{eqnarray*}}\hfill
   \parbox[b]{1.5cm}{\vspace{-0.0cm}\begin{eqnarrayabc}#2\end{eqnarrayabc}}\newline}
\ncm{\eqbox}[2]{\newline\parbox{1.5cm}{#1}\hfill
  \parbox{12cm}{\beanon #2\eeanon}\hfill
  \parbox{1cm}{\bea\eea}\newline}
\ncm{\nr}[1]{\parbox{1cm}{\begin{eqnarrayabc}#1\end{eqnarrayabc}}\\}

\ncm{\kal}[1]{\mbox{$\cal #1 $}}
\ncm{\mrk}[1]{\!\!\! #1 \!\!\!} 
\ncm{\qed}{\hspace*{0.4cm}\rule{0.24cm}{0.24cm}}  
\ncm{\mbold}[1]{\mbox{\boldmath $ #1 $}}   
\ncm{\bm}{\mbold}
\ncm{\str}{\stackrel}
\ncm{\sub}{\subset}
\ncm{\e}{\varepsilon}
\ncm{\ka}{\kappa}
\ncm{\inputc}[1]{\begin{center}\input{#1}\end{center}}
\ncm{\lto}{\longrightarrow}
\ncm{\x}{\times}
\ncm{\bmm}{\bm{\cal M}}
\ncm{\cp}{{\bf P}}    
\ncm{\bfp}{{\bf P}}
\ncm{\bmi}{\bm{i}}
\ncm{\bmom}{\bm{\om}}
\ncm{\bmOm}{\bm{\Om}}
\ncm{\res}{\restriction}
\ncm{\bmL}{\bm{\cal L}}
\ncm{\bmell}{\bm{\ell}}
\ncm{\bmE}{\bm{\cal E}}
\ncm{\bme}{\bm{e}}
\ncm{\bmpi}{\bm{\pi}}
\ncm{\bmr}{\bm{r}}
\ncm{\bmsigma}{\bm{\sigma}}
\ncm{\wt}{\widetilde}

\newcommand{\beaa}{\begin{eqnarray}}
\newcommand{\eeaa}{\end{eqnarray}}

\begin{document}

\author{{\bf Oscar Bolina}\thanks{Supported by FAPESP under grant
97/14430-2. {\bf E-mail:} bolina@math.ucdavis.edu} \\
Department of Mathematics\\
University of California, Davis\\
Davis, CA 95616-8633 USA\\
}
\title{\vspace{-1in}
{\bf The Precessing Top}}
\date{}
\maketitle

\begin{flushleft}
{\large \bf 1. Introduction}
\end{flushleft}
When a symmetric top is set spinning with angular velocity $\omega_{c}$
about its own axis, it traces out a circle about a vertical
direction with angular velocity $\omega_{z}$, as shown in Fig. 1.
\newline
In most textbooks \cite{{HRW},{S}, {Sk}} the {\it precessional} 
angular velocity $\omega_{z}$ is calculated under the assumption
that it is much  smaller than the {\it spin} angular velocity
$\omega_{c}$. This makes for a simplification of the calculation
by considering the total angular momentum as due to the spin
motion only, and thus directed along the symmetry axis.
\newline
In this case, the angular momentum along the symmetry axis is 
$L=I \omega_{c}$, where $I$ is the moment of inertia about this 
same axis. The tip of this vector describes a circle of radius 
$L\sin{\theta}$ around the vertical direction. The torque of 
gravity is $mgl\sin\theta$, where {\it m} is the mass of the top 
and {\it l} is the distance of its center of mass {\it C} to {\it O}. 
In a time interval $\Delta t$ the torque changes the angular momentum 
by $L{\sin\theta} \omega_{z} \Delta t$ in a direction tangent to the
circle. Equating torque to time change in angular momentum yields 
the usual formula $\omega_{z}={mgl}/{I\omega_{c}}$ for slow precession
\cite{S}.
\newline
In this note I drop the simplifying assumption that the total 
angular momentum is solely along the symmetry axis in order to 
obtain the following general expression for $\omega_{z}$,
\beq\label{omega}\label{om}
\omega_{z}=\frac{I\omega_{c} \pm \sqrt{I^{2}
\omega_{c}^{2}+4(I-I_{n})mgl\cos\theta}}
{2 (I_{n}-I)\cos\theta}~,
\eeq
where $I_{n}$ is the moment of inertia of the top about any axis 
normal to the symmetry axis at {\it O}.
\newline
There are two values for $\omega_{z}$ if
$I^{2}\omega^{2}_{c}+4(I-I_{n})mgl\cos\theta > 0$.
This condition is always satisfied 
\newline
when $I > I_{n}$. 
When $I < I_{n}$ it determines the minimum velocity
$\omega_{c}=(2/I) \sqrt{(I_{n}-I)mgl\cos\theta}$
\newline
for a steady precession to occur. 
\newline
Formula (\ref{omega}) is most useful for practical purposes when
$\omega_{c}$ is very large, in which case the appro\-ximation
$\sqrt{1+x}=1+x/2$ for the square root in (\ref{omega}) when
$x={4(I-I_{n})mgl\cos\theta}/{I^{2}\omega^{2}_{c}}$ is very small 
gives
\beq
\omega_{z}=\frac{I\omega_{c}}{2 (I_{n}-I)\cos\theta} \pm
\left ( \frac{I\omega_{c}}{2 (I_{n}-I)\cos\theta} -
\frac{mgl}{I\omega_{c}} \right ).
\eeq
This approximation for $\omega_{z}$ yields not only the previous
formula for slow precession $\omega^{(-)}_{z}={mgl}/{I\omega_{c}}$, 
but also the formula for {\it fast} precession
$\omega^{(+)}_{z}={I\omega_{c}}/{(I_{n}-I)\cos\theta}$.
\newline
In this last case, the sense of precession depends on whether 
$I < I_{n}$ or $I > I_{n}$. For the top shown in Fig. 1, and
for most ordinarily shaped child's top, $I_{n} > I$. 
\newline
Which precession occurs depends on how the top is set in motion. Once 
it is started at an angle $\theta$ to the vertical with angular velocity
$\omega_{c}$ and either of the above values for $\omega_{z}$, it
will continue to precess steadily.

\begin{flushleft}
{\large \bf 2. Analysis}
\end{flushleft}
The analysis that leads to (\ref{omega}) is elementary and amounts
to taking into account the angular momentum about the vertical 
direction. The angular velocity $\omega_{z}$ has components
$\omega_{z} \cos\theta$ along the symmetry axis, and $\omega_{z}
\sin\theta$ along an axis normal to it, as shown in Fig. 2. Thus
the angular momentum about the vertical direction can be decomposed
into a component $I\omega_{z} \cos\theta$ along the symmetry axis, and
a component $I_{n} \omega_{z} \sin\theta$ normal to the symmetry axis.
Since these two components also describe a circle around the vertical
direction, the previous simplified analysis can be applied separately 
here for the components along and normal the symmetry axis. Thus the 
correction for the change in total angular momentum in a time interval 
$\Delta t$ has now two steps.
\begin{itemize}
\item[a.] The change in the component of the angular momentum
along the symmetry axis is what we had before, with the addition 
of the extra term $I \omega_{z}\cos\theta$ due to the 
precession motion. It becomes $(L+I\omega_{z}
{\cos\theta})\sin\theta \omega_{z} \Delta t$. 

\item[b.] The component of the angular momentum normal to the
symmetry axis is $L_{n}=I_{n} \omega_{z} \sin{\theta}$. The tip 
of this vector describes a circle of radius $L_{n}\cos{\theta}$ 
around the vertical direction. The torque of gravity changes this
component in a direction tangent to this circle. Thus the change
in the normal component is given by  $L_{n}\cos{\theta}\omega_{z} 
\Delta t$. 
\end{itemize}
Since the changes in {\it (a)} and {\it (b)} are opposite 
in direction, equating torque to time change in angular 
momentum now leads to the equation
$(I-I_{n})\omega^{2}_{z}\cos\theta  
+(I\omega_{c})\omega_{z}-mgl=0$ for $\omega_{z}$, whose
solutions are given by (\ref{omega}).
\newpage
\noindent


\newpage
\begin{figure}
\centerline{
\epsfbox{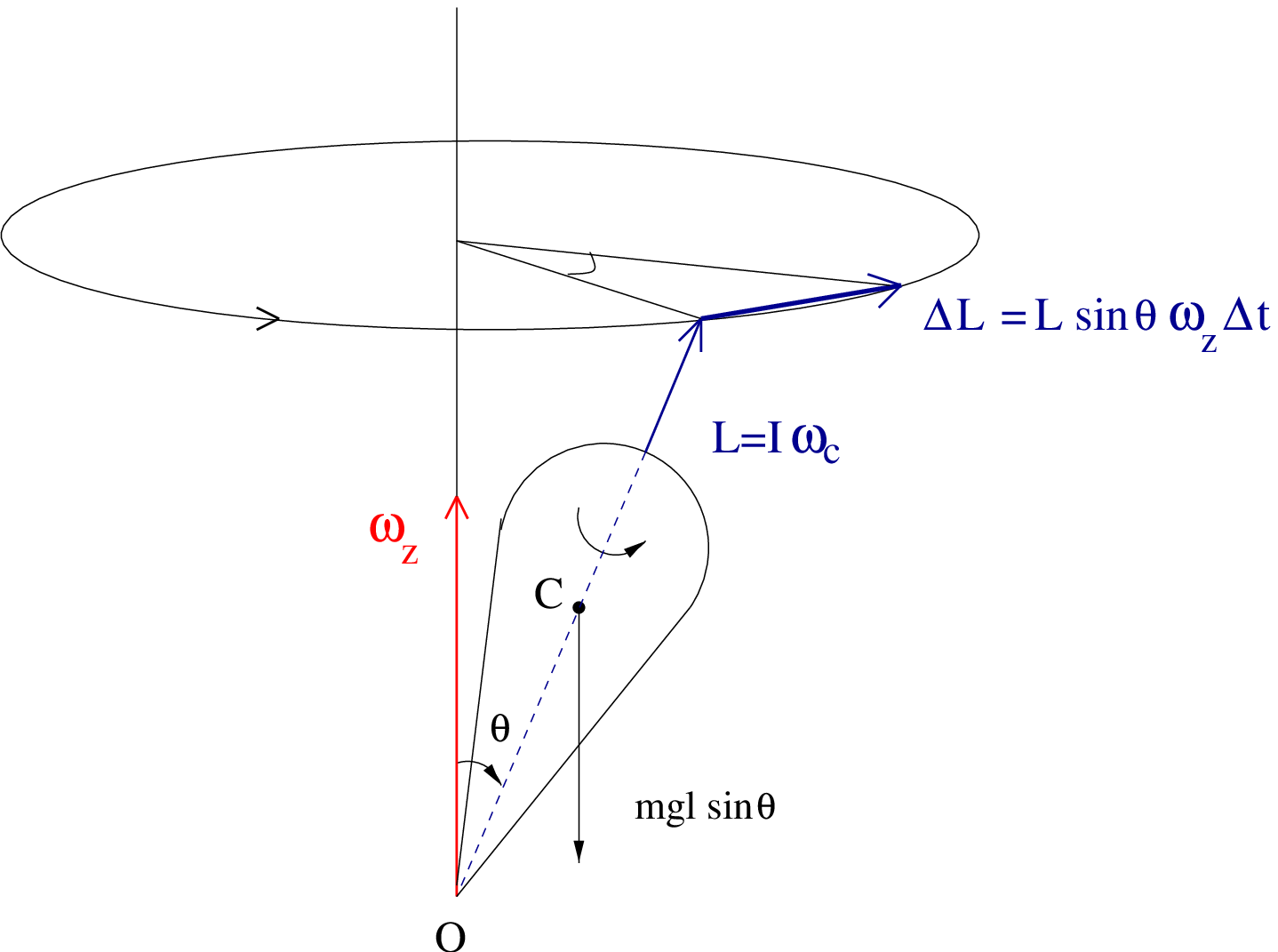}}
\vskip 5 cm
\noindent
\caption{In the simplified analysis the total angular momentum
of the top $L=I\omega_{c}$ is due solely to the rotation 
around its axis of symmetry.}
\end{figure}
\newpage
\begin{figure}
\centerline{
\epsfbox{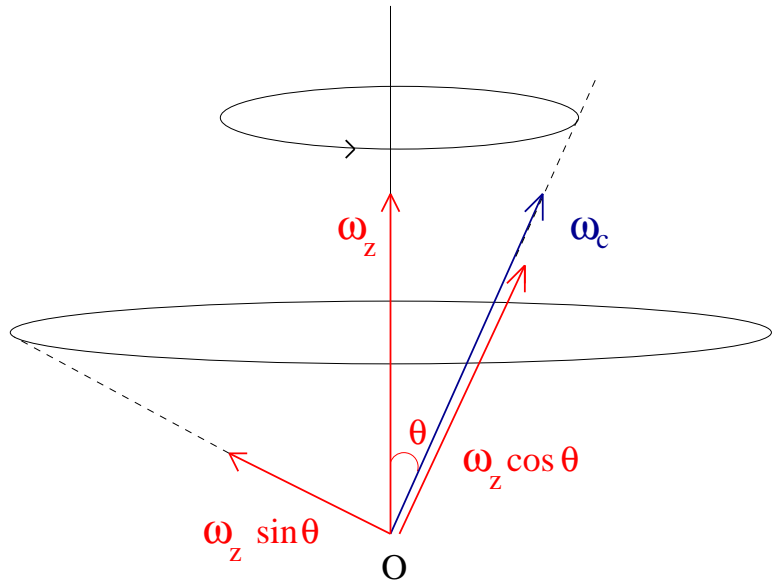}}
\vskip 5 cm
\noindent
\caption{In the general analysis, the angular momentum of 
the rotational motion around the vertical axis is taken into account.}
\end{figure}

\end{document}